# III-V quantum light source and cavity-QED on Silicon

I. J. Luxmoore[1*], R. Toro[1], O. Del Pozo-Zamudio[1], N. A. Wasley[1], E. A. Chekhovich[1], A. M. Sanchez[2], R. Beanland[2], A. M. Fox[1], M. S. Skolnick[1], H. Y. Liu[3] and A. I. Tartakovskii[1]

[1]*Department of Physics and Astronomy, University of Sheffield, Sheffield, S3 7RH, UK*

[2]*Department of Physics, University of Warwick, Coventry CV4 7AL, UK33*

[3]*Department of Electronic and Electrical Engineering, University College London, London, WC1E 7JE, UK*

[*]*email: i_luxmoore@hotmail.com*

**Non-classical light sources offer a myriad of possibilities in fundamental science and applications including quantum cryptography[1] and quantum lithography[2]. Single photons can encode quantum information and multi-qubit gates in silica waveguide circuits[3] have been used to demonstrate linear optical quantum computing[4]. Scale-up requires miniaturisation of the waveguide circuit and multiple photon sources. Silicon photonics, driven by the incentive of optical interconnects[5], is a highly promising platform for the passive components[6], but integrated light sources are limited by silicon's indirect band-gap. III-V semiconductor quantum-dots, on the other hand, are proven quantum emitters[7–11]. Here we demonstrate single-photon emission from quantum-dots coupled to photonic crystal nanocavities fabricated from III-V material grown directly on silicon substrates. The high quality of the III-V material and photonic structures is emphasized by observation of the strong-coupling regime. This work opens-up the advantages of silicon photonics to the integration and scale-up of solid-state quantum optical systems.**

Incorporating photonic components onto a Si platform has been a powerful driver behind the development of Si photonics for the last twenty years, with a key motivation being the development of on-chip and chip-chip optical interconnects[5] However, the indirect band-gap of Si has severely restricted the development of light sources integrated directly with Si components. III-V semiconductor materials provide a mature technology with light emitting devices across a wide spectrum from the UV to the mid infra-red, but the direct growth on Si is difficult because of lattice mismatch. Despite these difficulties, there have been several demonstrations of light emitting devices grown on Si substrates in recent years, including GaN quantum well[12] and InGaAs/GaAs quantum-dot (QD) lasers monolithically grown on Si[13,14] and with Ge virtual substrates[15]. A Ge virtual substrates was also used recently for GaAs QDs grown by droplet epitaxy[16]. Integrating quantum light sources with Si can realise new circuit functionality as well as, in the long-term, reduce the production costs of QD quantum light sources for commercial applications such as quantum key distribution[1].

InGaAs/GaAs QDs have been widely employed in non-classical light sources[7–9]. Indistinguishable photon emission from two remote QDs[10,11] has opened-up the possibility of transferring quantum information between remote solid-state systems and QDs can emit polarization-entangled photons with near unity probability via the two-photon cascade of the biexciton state[8,9]. QDs are also highly compatible with photonic structures, such as micropillars[9,17–19] and photonic crystal cavities[20–22], which can be exploited to maximise the collection efficiency[9], enhance the spontaneous emission rate[18,23] and enter the regime of strong light-matter coupling[17,19–22].

In this work, we demonstrate that high quality and low density InGaAs QDs can be grown directly on a Si substrate. Photonic crystal cavities are fabricated using this material and employed to enhance the single-photon emission rate and collection efficiency, thus demonstrating the potential for the integration of a high-efficiency, deterministic single-photon source with Si photonics. The high quality of the material has enabled fabrication of photonic crystal cavities with Q-factors exceeding 13,000. Furthermore, the strong-coupling regime of a QD and the optical field of a nanocavity is observed: characteristic anti-crossing behaviour with a Rabi splitting of 212μeV is measured in photoluminescence (PL) measurements by tuning the sample temperature.

The wafers studied in this work are grown using molecular beam epitaxy with the layer structure shown in Fig. 1(a) (see Methods for further details). A 1μm layer of GaAs is grown monolithically on a silicon wafer. Due to the high lattice mismatch of 4.2%, threading dislocations nucleate at the interface and propagate through the GaAs layer, leaving surface defects. In order to reduce the density of these dislocations a strained layer superlattice (SLS) is grown using 5 alternating layers of $In_{0.15}Al_{0.85}As$/GaAs (10nm/10nm), capped with 300nm of GaAs and repeated 4 times. To obtain the very smooth surface required for low density QD growth and high quality photonic structures, a short period superlattice (SPL) consisting of 50 alternating layers of $Al_{0.4}Ga_{0.6}As$/GaAs (2nm/2nm) is then grown and capped by 300nm GaAs[14]. Following this, the 1μm $Al_{0.6}Ga_{0.4}As$ sacrificial layer, required for photonic crystal fabrication is grown and a 140nm GaAs layer containing the InGaAs QDs at its centre completes the structure. The nominally InAs QDs are grown at 500°C using a growth rate of 0.016ML/s with the QD emission energy controlled using the In-flush technique[24]. The cross-sectional transmission electron microscope image shown in Fig. 1(d) highlights the effectiveness of the dislocation filter layers in reducing the defect density, which is measured using etch-pit density measurements to be ~$6\times10^6$cm$^{-2}$ in the GaAs layer directly above the SPL.

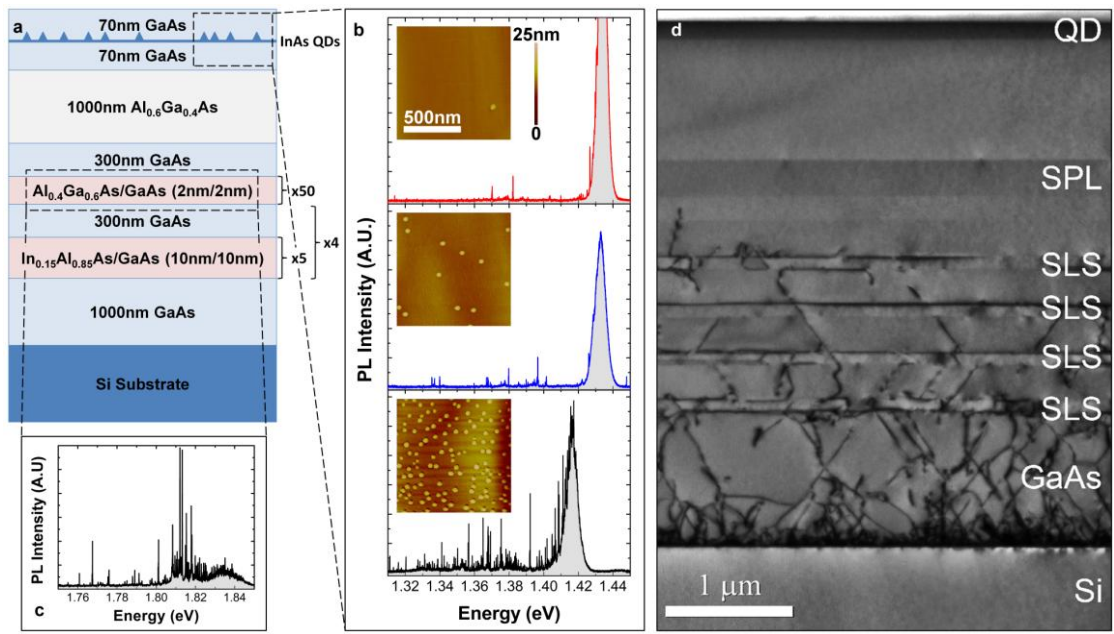

**Figure 1.** Layer structure of the QD on Si wafers and material characterisation. (a) Schematic diagram showing the III-V semiconductor layers grown on the Si substrate. (b) Photoluminescence spectra recorded from the InGaAs/GaAs QDs at different locations on the three inch as-grown wafer. The intense, broad peak between 1.41 and 1.43eV corresponds to emission from the QD wetting layer. Narrow spectral lines originating from charged and neutral exciton complexes within individual QDs are observed in the range of 1.3–1.4eV. The insets show AFM scans from the corresponding locations on an un-capped wafer. (c) Photoluminescence spectra showing narrow emission lines which originate from the AlGaAs/GaAs short-period superlattice. (d) Cross-sectional bright-field transmission electron microscope image of the as-grown wafer, highlighting the capture of threading dislocations by the strained layer superlattices and the short period superlattice structures.

To assess the quality of the material we use atomic force microscopy (AFM) and PL spectroscopy (see Methods). Fig. 1(b) shows PL spectra recorded from different positions on the wafer and corresponding AFM images of an uncapped sample. The spectra are as expected for InGaAs/GaAs QDs, with a broad peak at 1.43eV corresponding to emission from the QD wetting layer. Narrow spectral lines originating from charged and neutral exciton complexes within individual QDs are observed in the range of 1.3–1.4eV and have resolution-limited linewidths of ~30μeV. Due to a small variation in temperature across the 3 inch diameter wafer during growth the QD density varies between $1\times10^8$ and $1.5\times10^{10}$cm$^{-2}$, as illustrated by the PL measurements and AFM images, shown in Fig. 1(b). QD-like emission is also observed at higher energy, in the range 1.75-1.85eV, an example of which is shown Fig. 1(c). PL measurements of samples etched to different depths reveal the origin of this emission to be the AlGaAs/GaAs superlattice. The formation and optical properties of QDs in this superlattice is currently the subject of further investigations.

To demonstrate the potential of the InGaAs QDs grown on Si as a single-photon source, photonic crystal cavities are fabricated using an area of the sample with moderate QD density of 5-10$\times10^9$cm$^{-2}$ (see Methods). Fig. 2(a) shows a scanning electron microscope image of an L3-defect photonic crystal cavity. The maximum Q-factor measured in these devices is ~13,000. This is similar to the maximum value measured for devices fabricated with the same process but on GaAs substrates, suggesting that the optical quality of the GaAs is high, despite the Si substrate.

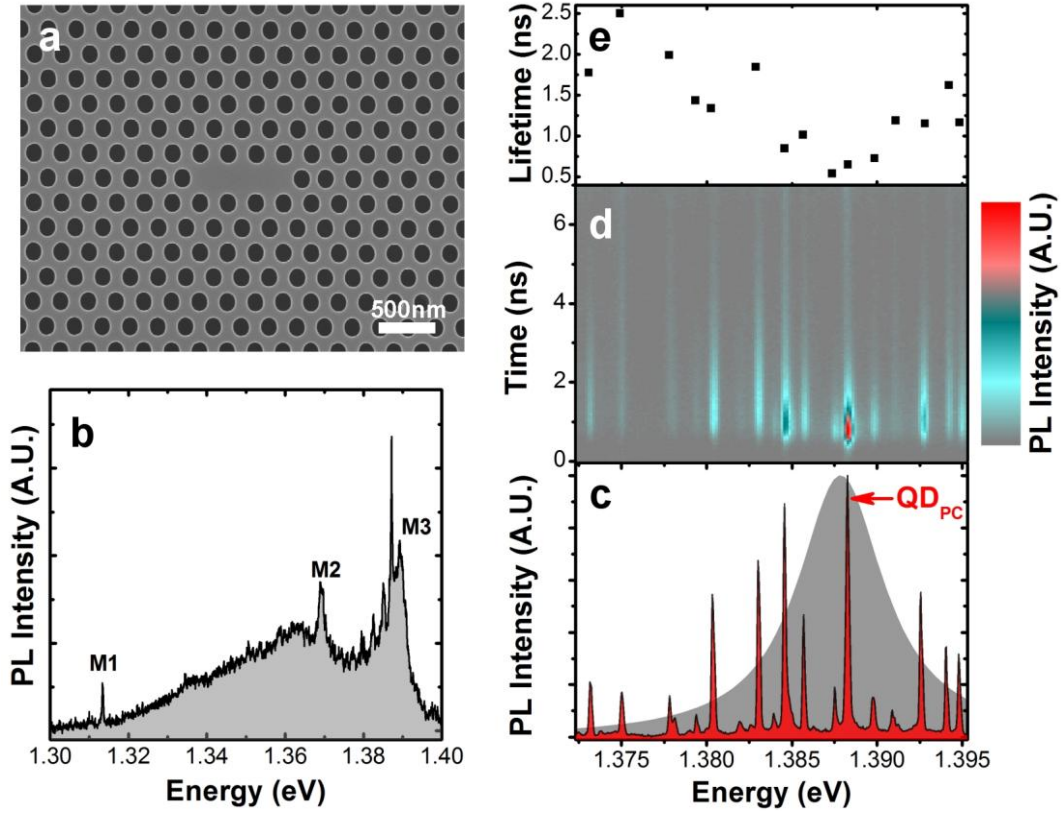

**Figure 2.** Photonic crystal cavity QD single-photon source. (a) Scanning electron microscope image showing a typical L3 defect photonic crystal cavity. (b) High power PL spectrum of the L3 cavity. (c) PL spectra recorded at low excitation power, showing emission of QDs coupled to the photonic crystal cavity mode, M3. The grey shaded area represents the spectral location and linewidth of the cavity mode. (d) Time resolved PL measurements recorded as a function of emission energy for the QD-photonic crystal cavity. (e) Lifetimes extracted from time-resolved PL measurements shown in (d) as a function of emission energy.

To demonstrate single-photon emission, the photonic crystal cavity is employed to enhance the spontaneous emission rate and increase the extraction efficiency of the QD. In this case, we employ not the fundamental cavity mode, but one of the higher order modes which has greater out of plane leakage[25]. We use the third lowest energy mode, M3 as indicated in Fig. 2(b), which has a Q-factor of ~250. The motivation for using this mode is two-fold; firstly, the large out-of-plane leakage means that the emission can be more efficiently collected into the microscope objective and secondly, the low Q-factor greatly increases the likelihood of finding spectral overlap between a QD and the cavity resonance. The main drawback is that the low Q-factor restricts the maximum Purcell factor, $F_p$ that can be achieved; however, the small mode volume of the cavity means that a reasonably large $F_p$ up to ~30 can still be expected. At low excitation power, the cavity mode spectrum, shown in Fig. 2(c) reveals several bright lines corresponding to individual QDs in, and close to, resonance with the cavity mode.

To investigate the coupling between the QDs and cavity mode, time-resolved PL measurements are performed as a function of emission energy, as presented in Fig. 2(d). From exponential fits to this data the QD lifetime is extracted and plotted in Fig. 2(e). Although the exact spatial position of each QD within the cavity is different, there is a clear reduction in the lifetime as the spectral detuning between the QD and mode decreases, with a minimum measured lifetime of 0.54ns for the QD at 1.3875eV, where the resolution of the measurement system is ~350ps.

To characterise the performance of the QD single-photon source, we compare the QD line at ~1.388eV, labeled $QD_{PC}$ in Fig. 2(c), with a typical QD in the bulk material, away from the photonic crystal, $QD_{bulk}$. Fig. 3(a) shows the power dependent PL intensity of $QD_{PC}$, compared with the exciton (X) and biexciton (XX) power dependence of $QD_{bulk}$. For $QD_{bulk}$ the intensity follows the linear (quadratic) behavior consistent with that of the X(XX) in InGaAs QDs and saturates at an excitation power of ~1μW. Similarly, $QD_{PC}$ displays a linear dependence on excitation power and saturates at a similar excitation power, but at a PL intensity ~54 times brighter, consistent with the enhanced collection efficiency afforded by the photonic crystal cavity[25]. Fig. 3(b) compares the time-resolved PL of $QD_{PC}$ and the X transition of $QD_{bulk}$ and shows a significantly shorter lifetime of 0.64ns compared with 1.1ns for $QD_{bulk}$ (average lifetime for QDs in the bulk is 1.22±0.18ns), which corresponds to a Purcell enhancement of ~2 and confirms the regime of weak-coupling between the QD exciton and the cavity mode.

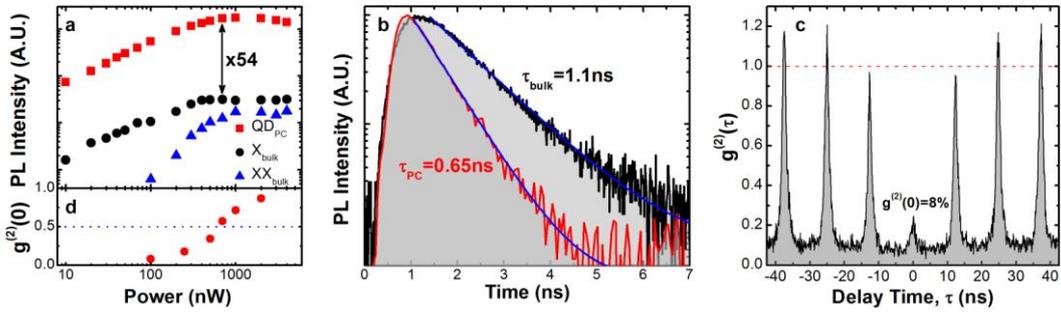

**Figure 3.** Performance of cavity coupled QD single-photon source. (a) Integrated PL intensity of a cavity-coupled QD compared to that of the exciton (X) and biexciton (XX) of a typical QD in the bulk as a function of pulsed-laser excitation power. (b) Time-resolved measurements of the lifetime measured for the bulk neutral exciton and cavity-coupled QDs shown in (a). (c) Photon coincidence histogram recorded from the cavity-coupled QD. (d) Normalised area of central peak of second order correlation function, $g^{(2)}(0)$, as a function of excitation laser power. The dashed line shows the limit of single-photon emission at $g^{(2)}(0)=0.5$.

The single-photon emission is investigated with a Hanbury-Brown Twiss measurement. Fig. 3(c) plots the $g^{(2)}(\tau)$ function recorded from $QD_{PC}$ at a pulsed excitation power of 100nW. Clear anti-bunching is observed with a multi-photon emission probability $g^{(2)}(0)=0.08$, demonstrating the single-photon nature of the emission. To determine the maximum single-photon emission rate, $g^{(2)}(\tau)$ is measured for different excitation powers. Fig. 3(d) plots the multi-photon emission probability, $g^{(2)}(0)$ as a function of the excitation power and reveals a monotonic increase with power, as the single-photon emission from the QD saturates but the background cavity emission continues to increase[19]. At the saturation power of 500nW the single-photon detection rate is ~80kHz, with $g^{(2)}(0)=0.4$.

In a second structure, we observe the regime of strong-coupling between the optical field of the cavity mode and a single QD. In this case the fundamental mode of the L3 cavity has a Q-factor of ~8,000. At the base temperature of ~10K, the QD is blue detuned from the cavity mode by 830μeV. By increasing the sample temperature the QD can be tuned into resonance with the cavity mode as shown in Fig. 4(a). As the QD is tuned through the mode resonance, two distinct peaks are observed in the spectra at all temperatures. This anti-crossing is the signature of the strong coupling regime, where there is a reversible exchange of energy between the QD and the cavity mode resulting in the vacuum Rabi splitting (VRS), which has been observed in several QD based systems[17,19–22].

Fig. 4(b) shows a plot of the peak energies extracted from the temperature dependent spectra. The complex energies of the upper and lower polariton branches, $E_\pm$, of the strongly coupled system can be described by the equation[26]

$$E_\pm = \frac{E_m + E_x}{2} - i\frac{\gamma_m + \gamma_x}{4} \pm \sqrt{g^2 + \frac{1}{4}\left(E_m - E_x + i\frac{\gamma_m - \gamma_x}{2}\right)^2},$$

(1)

where $E_m$ and $E_x$ are the energies of the uncoupled mode and QD exciton, respectively, $\gamma_m$ and $\gamma_x$ are the linewidths of the cavity mode and QD exciton, respectively and *g* is the coupling constant. The upper and lower polariton energies, calculated using Eq. (1) are plotted in Fig. 4(b) for a zero detuning VRS, 2g=212µeV, which shows good agreement with the experimental data. The linewidth of the cavity mode at ~10K is 174µeV, giving a ratio $g/\gamma_m = 0.61 > 1/4$, thus fulfilling the condition for strong-coupling[26]. The large VRS, ~75% of the predicted value[27], suggests that the degree of spatial overlap between the QD and cavity mode is high and compares well with values reported for similar systems[20–22].

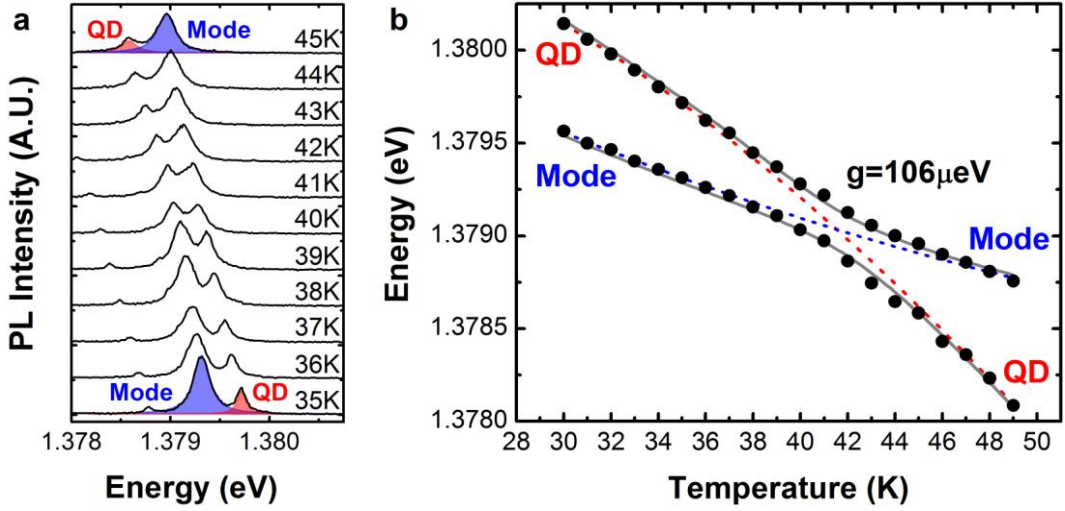

**Figure 4.** Strong light-matter coupling between a quantum dot and the photonic mode of an L3 nanocavity. (a) Temperature dependent photoluminescence spectra showing the anti-crossing of QD and cavity mode peaks. (b) Energies of the upper and lower polaritons, extracted from data shown in (a). The grey line shows a fit to the data of equation (1) and the dashed red and blue lines show estimates of temperature dependence of the uncoupled QD and cavity mode respectively.

In summary, we have presented the integration of high quality quantum emitters with a Si substrate. Numerous challenges remain, but the demonstration of the strong coupling regime proves that sophisticated III-V optical devices can be integrated with a Si platform. One such challenge is to develop an efficient means of coupling the quantum light into a Si photonic circuit. Whispering gallery sources laterally coupled to Si waveguides[28] provide one possible solution and epitaxial layer overgrowth[29] another. With the addition of on-chip superconducting single-photon detectors[30] all the elements required for scalable linear optical quantum computing can be combined in a single Si based platform.

## Methods

**Molecular Beam Epitaxy.** Phosphorus-doped (100)-orientated 3 inch Si substrates with 4° offcut towards the [110] planes are used in the experiments. Prior to growth, oxide desorption is performed by holding the Si substrate at a temperature of 900°C for 10 minutes. The Si substrate is then cooled down for the growth of a 30-nm GaAs nucleation layer with a low growth rate of 0.1ML/s, followed by the 1μm GaAs layer grown at high temperature with a higher growth rate.

**Photonic Crystal Fabrication.** Photoluminescence measurements are used to identify an area of the wafer with QD density of ~5x10$^9$cm$^{-2}$, which is then employed for the fabrication of the photonic crystal nanocavities. Electron beam lithography is used to define the photonic crystals and the GaAs slab layer is etched using a chlorine based inductively coupled plasma reactive ion etch (ICP-RIE). Finally, hydrofluoric acid is used to selectively remove the $Al_{0.6}Ga_{0.4}As$ layer from beneath the photonic crystals, leaving the free-standing photonic crystal membrane.

**Optical Measurements**

The optical measurements are performed in a liquid helium flow cryostat with a base temperature of ~10K. The PL is excited using a continuous wave (CW) or pulsed laser tuned to 850nm, focused to a ~1μm diameter using a 50X microscope objective (NA=0.42). For the time-resolved measurements, presented in Figures 2 and 3 a Ti:Sapphire laser with a pulsewidth of ~100fs is used to excite the PL. The emission from a single QD is filtered using a single grating spectrometer and detected with a charge coupled device (CCD) camera or avalanche photo-diode (APD), which has a time-resolution of ~350ps. In the case of the $g^{(2)}$ measurements the light filtered by the spectrometer is split by a fiber beam-splitter and coupled to a pair of APDs. For the strong-coupling measurements, presented in Fig. 4, a CW diode laser tuned to 850nm is used to excite the PL, which is dispersed using a double grating spectrometer and detected with a CCD camera.

## Acknowledgements

This work was supported by the EPSRC Programme grants (EP/G001642/1 and EP/J007544/1) and ITN Spin-Optronics. O.D.P.Z. was supported by a CONACYT-Mexico doctoral scholarship.